\documentclass[a4paper,11pt]{article}
\pdfoutput=1 

\usepackage{jcappub} 

\usepackage[T1]{fontenc} 
\usepackage{graphicx}
\newcommand{\hmpc}{{\, h^{-1}\, {\rm Mpc}}}

\def\aj{AJ}
\def\apj{ApJ}

\def\jcap{JCAP}
\def\mnras{MNRAS}
\def\aap{A\&A}

\def\nat{Nature}      

\def\apjl{ApJ Letters}
\def\physrep{Physics Reports}
\def\qcg{Classical and Quantum Gravity}

\title{\boldmath Testing homogeneity of the galaxy distribution in the SDSS using Renyi entropy}

\author[a]{Biswajit Pandey,}
\author[b]{Suman Sarkar}

\affiliation[a]{Department of Physics, Visva-Bharati University,
  Santiniketan, 731235, India} \emailAdd{biswap@visva-bharati.ac.in}
\affiliation[b]{Department of Physics, Indian Institute of Science
  Education and Research Tirupati, Tirupati - 517507.  Andhra Pradesh,
  India} \emailAdd{suman2reach@gmail.com}

\abstract{We analyze a set of volume limited sample of galaxies from
  the SDSS to study the issue of cosmic homogeneity. We use the Renyi
  entropy of different order to probe the inhomogeneties present in
  the galaxy distributions. We also calculate the Renyi diveregence to
  quantify the deviations of the galaxy distribution from a
  homogeneous Poisson distribution on different length scales. We
  separately carry out the analysis using the overlapping spheres and
  the independent voxels. Our analysis suggests that the scale of
  homogeneity is underestimated in the smaller galaxy samples due to
  the suppression of inhomogeneities by the overlapping of the
  measuring speheres. We find that an analysis with the independent
  voxels and/or use of a significantly larger galaxy sample can help
  to circumvent or mitigate this problem. Combining the results from
  these analyses, we find that the galaxy distribution in the SDSS
  becomes homogeneous on a length scale beyond $140 \hmpc$.}

\begin{document}
\maketitle
\flushbottom

\section{Introduction}
The statistical homogeneity and isotropy on large scales is a
fundamental prerequisite to our current understanding of the Universe.
The standard cosmological model heavily relies on this assumption. The
assumption is not provable in a mathematical sense and can be only
tested against various cosmological observations. The galaxies are the
basic units of the large-scale structures in the Universe. The
distribution of galaxies trace the large-scale distribution of matter
in the Universe. The present generation galaxy surveys
\citep{colless,york} reveal that the galaxies are embedded in a highly
inhomogeneous complex weblike network of filaments, sheets and
clusters surrounded by nearly empty regions. It is important to test
if the galaxy distribution exhibit a transition to homogeneity on
sufficiently large-scales.

The homogeneous and isotropic FRW space-time geometry provides great
simplicity and ease in describing the global structure of the
Universe. In modern cosmology, the FRW geometry is indispensable for
analysis and interpretation of data from the cosmological
observations. The FRW geometry is a mathematical idealization and
requires validation from observations. The present day observable
Universe is believed to have originated from the tiny primordial
inhomogeneities imprinted on the CMBR maps. The inhomogeneous
distribution of matter in the present Universe is a product of the
amplification of these seed inhomogeneities by gravity. The
inhomogeneity of the matter distribution is thus inherent and are
expected to be present upto certain length scales. However, if the
inhomogeneities persist on the largest length scales then it would
have several implications for cosmology. The inhomogeneities, by
virtue of their backreaction on the metric may provide an alternate
explanation for the observed cosmic acceleration \citep{buchert97,
  schwarz, kolb06, buchert08, ellis}. The inhomogeneous cosmological
models have been explored in a number of comprehensive reviews
\citep{marranotari, bolejko, redlich}.

Many observations such as the CMBR \citep{penzias,smoot,fixsen}, radio
sources \citep{wilson,blake}, X-ray background
\citep{peeb93,wu,scharf}, Gamma ray bursts \citep{meegan,briggs},
supernovae \citep{gupta,lin} and galaxies
\citep{marinoni,alonso,sarkariso19} support the isotropy of the
Universe. The homogeneity and isotropy of the Universe are two
different aspects which may or may not simultaneously hold for our
Universe. So it is important to validate the assumption of homogeneity
using various cosmological observations. The homogeneity of the
Universe is comparatively more difficult to confirm.  This is caused
by the difficulty in distinguishing the spatial variation from
temporal evolution in the past light cone and the redshift-dependent
selection effects in the sample. So testing the statistical
homogeneity of the Universe on very large-scales is a challenging
task. One may minimize these effects by focusing only on the galaxy
distribution in the nearby Universe which may be approximately treated
as a constant time spatial hypersurface.

The study of homogeneity with galaxy distribution has a long history
dating back to 80s and 90s when the spectroscopic data were made
available by first generation galaxy redshift surveys. Early studies
show that the galaxy distributions exhibit a fractal nature with a
scale invariant behaviour on small scales
\citep{pietronero,coleman92,mandelbrot}. Such scale-invariant feature
of the galaxy distribution are shown to exist even on large scales by
a number of studies \citep{pietronero,coleman92,amen,joyce,labini07,
  labini09, labini11}. The existence of the fractal nature of the
galaxy distribution out to the scale of the survey is clearly in
disagreement with the assumption of homogeneity. But many other
studies, though confirm the fractal nature of the galaxy distribution
on small scales, reported that the galaxy distribution is homogeneous
on scales $70-150 \hmpc$
\citep{martinez94,borgani95,guzzo97,cappi,bharad99,pan2000,yadav,hogg,
prakash,scrim,nadathur,pandeysarkar15,pandeysarkar16}.

The present day Universe exhibits a hierarchy of structures where
galaxies are assembled into groups and clusters which are then
interwoven into larger filamentary superclusters. Gott et
al. \citep{gott05} identified the Sloan Great Wall (SGW) in the SDSS
which nearly extends upto $\sim 400$ Mpc. Clowes et al. \citep{clowes}
reported the existence of a large quasar group (LQG) spanning more
than $500 \hmpc$ at $z\sim 1.3$.  Observations suggest that there
exists enormous empty regions such as the Eridanus void stretching
across $\sim 300$ Mpc \citep{szapudi} and the KBC void which is a
nearly spherical void with diameter of $\sim 600$ Mpc
\citep{kbc13}. Existence of such coherent large-scale structures may
pose a challenge to the assumptions of homogeneity and isotropy and
the statistical significance of any such structures must be assessed
carefully \citep{park12, nadathur}.

The multi-fractal analysis of the galaxy distribution
\citep{martinez90, coleman92, borgani95, bharad99, yadav} remains one
of the most popular method for the study of homogeneity. It is based
on the scaling of different moments of the number counts of galaxies
in spheres centered around the galaxies. The Renyi dimension
\citep{renyi70} or the generalized dimension \citep{hentschel} can be
used to characterize multi-fractals. A mono-fractal can be considered
as homogeneous when it has the same generalized dimension for
different moments and all these generalized dimensions coincide with
the ambient dimension. But an accurate measurement of generalized
dimension or Renyi dimension are hard to achieve due to the finite and
discrete nature of the galaxy distributions and the $r \to 0$ limit in
these definitions \citep{saslaw99}. Pandey \citep{pandey13} propose an
alternative statistical measure based on the count-in-spheres
statistics and Shannon entropy \citep{shannon48} and use it to measure
the scale of homogeneity in the Main Galaxy sample
\citep{pandeysarkar15}, LRG sample \citep{pandeysarkar16} and quasar
sample \citep{sarkarpandey16} from the SDSS. Renyi \citep{renyi61}
provided a more generalized concept of entropy which can be used to
quantify the uncertainty or randomness of a system. The Renyi entropy
includes the Shannon entropy as a special case. The Shannon entropy is
represented by the Renyi entropy of order one. The Renyi entropies of
higher order are increasingly determined by the events with higher
probability. Naturally, they are more sensitive to the inhomogeneities
present in a distribution. Recently, Pandey \citep{pandey21} defined a
measure of homogeneity based on the Renyi entropy.

The Sloan Digital Sky Survey (SDSS) is the most successful redshift
survey of all times. It has measured the photometric and spectroscopic
information of millions of galaxies, which revolutionized our
knowledge about the large-scale structures in the Universe. The high
quality and volume of data provided by the SDSS has led to us to an
era where many important cosmological questions can be addressed in a
manner which were never possible before. The SDSS has reveled the
three-dimensional distribution of galaxies in the nearby Universe in
its full glory.

In the present work, we employ a Renyi entropy based measure
\citep{pandey21} to study the issue of homogeneity using the SDSS main
galaxy sample. We compare our findings against the results obtained
from the previous studies.

The plan of the paper is the following. We outline our method in
Section 2 and describe the data in section 3. The results and
conclusions of the analysis are presented in Section 4.

\section{Method of Analysis}
The Renyi entropy \citep{renyi61} provides a more generalized concept
of entropy which was originally proposed by Alfred Renyi. The Renyi
entropy associated with a discrete random variable $X$ is given by,\\
\begin{eqnarray}
S_q(X) & = & \frac{1}{1-q}\, \log \sum^{n}_{i=1} \, p^q(x_{i})
\label{eq:two}
\end{eqnarray}
, where $p(x_i)$ is the probability of $i^{th}$ outcome and $q \in
    [0,\infty]$. We have $\{x_{i}:i=1,....n\}$ with a total $n$
    outcomes. It can be shown that for $q=1$, the Renyi entropy is
    same as the Shannon entropy. The Renyi entropy $S_q$ is a mildly
    decreasing function of $q$. For higher values of $q$, the Renyi
    entropy is increasingly decided by the the events with higher
    probabilities. So the higher order Renyi entropies are more
    sensitive to inhomogeneitis as compared to the Shannon entropy.
    If probabilities of all the events become equal then we have
    $S_q(X)=\log n$ independent of the order $q$ of the Renyi entropy.
    Pandey \citep{pandey21} propose a simple measure of homogeneity
    for galaxy distribution based on this property of Renyi entropy.

The Renyi dimension or the generalized dimension $d_q$ \citep{renyi70}
is a widely used measure of homogeneity. The Renyi dimension of order
$q$ is defined as,\\
\begin{eqnarray}
d_q(X) & = & \lim_{\epsilon \to 0} \frac{S_q(X)}{\log \frac{1}{\epsilon}} 
\label{eq:rendim}
\end{eqnarray}
Here $\epsilon$ is the scaling factor. The $\epsilon \to 0$ limit in
this definition can not be achieved in a meaningful way in galaxy
samples with finite number of galaxies. This often leads to
inaccuracies in the measured spectrum of the generalized
dimension. The Renyi dimension is estimated for both the positive and
the negative values of $q$. The positive and the negative values of
$q$ assign greater weights to the overdense and the underdense regions
respectively. In a finite sample, the generalized dimension for the
negative values of $q$ are very sensitive to the underdensities
present in the sample \citep{roberts}. This may dramatically affect
the generalized dimension for $q<0$ making it difficult to distinguish
empty space from regions with large underdensities. The method used in
the present analysis is based only on the maximization of
uncertainty. It does not require any such limit and is restricted to
only positive values of $q$. The generalized dimension is also prone
to the effects of survey geometry and incompleteness \citep{scrim,
  avila} whereas the normalized entropies are less susceptible to
these issues \citep{pandeysarkar15}.

    Given a distribution of $N$ galaxies over a volume $V$, we
    consider a sphere of radius $r$ centered around each galaxies and
    count the number of galaxies $n(<r)$ within it. The number count
    around the $i^{th}$ galaxy is given by,\\
\begin{eqnarray}
 n_i(<r)=\sum_{j=1}^{N}\Theta(r-\mid {\bf{x}}_i-\bf{x}_j \mid)
\label{eq:four}
\end{eqnarray}
, where $\Theta$ is the Heaviside step function. Here ${\bf{x}_{i}}$
and ${\bf{x}_{j}}$ are the radius vector corresponding to the $i^{th}$
and $j^{th}$ galaxies respectively. The number counts for the galaxies
near the boundary of the volume will be smaller due to partial
coverage. We take into account this effect by ignoring all the centers
which lie within a distance $r$ from the boundary of the volume. This
provides us with a finite number of centers $M(r)$ for each specific
radius $r$. The number of available centres will decrease with the
increasing radius due to the finite volume occupied by the
distribution. These $M(r)$ centres at each $r$ can be used to
calculate the Renyi entropies of the distribution. We can define a
random variable $X_r$ corresponding to each radius $r$. If a center is
randomly chosen from the $M(r)$ centers available at a radius $r$ then
there are a total $M_r$ possible outcomes. The probability of choosing
the $i^{th}$ centre is $f_{i,r}=\frac{\rho_{i,r}}{\sum^{M(r)}_{i=1} \,
  \rho_{i,r}}$. This is decided by the density at the location of the
$i^{th}$ centre which is given by
$\rho_{i,r}=\frac{n_{i}(<r)}{\frac{4}{3}\pi r^{3}}$. By definition, we
have, $\sum^{M(r)}_{i=1} \, f_{i,r}=1$. We can then define the Renyi
entropy of order $q$ associated with the random variable $X_{r}$ as,\\
\begin{eqnarray}
S_{q}(r) & = & \frac{1}{1-q} \log \sum_{i=1}^{M(r)} \, f^q_{i,r} \nonumber\\
& = & \frac{1}{1-q} \log \frac{\sum_{i=1}^{M(r)} n_i^q(<r)}{(\sum^{M(r)}_{i=1} n_i(<r))^q} 
\label{eq:five}
\end{eqnarray}\\
The base of the logarithm is arbitrary and here we choose it to $10$.
If the distribution is perfectly homogeneous on a length scale $r$,
then the spheres centered on $M(r)$ centres available at that radius
are expected to contain exactly same number of galaxies within them.
It implies that $f_{i,r}=\frac{1}{M(r)}$ for each centre i.e. all
centres are equally likely to be selected. This maximizes the
uncertainty in $X_r$. All the Renyi entropies of different orders
would reduce to $S_q(r)=\log \,M(r)$ under such a situation. We label
this maximum entropy as $[S_q(r)]_{max}$. We calculate the Renyi
entropies of different orders at each length scale $r$ and then
normalize it by the maximum entropy as $[S_q(r)]_{max}$ corresponding
to that radius. When the distribution is homogeneous the ratio
$\frac{S_q(r)}{[S_q(r)]_{max}}$ for different $q$ values are expected
to be $1$ or in other words, all the Renyi entropies of different
orders have the same value. In a real situation,
$\frac{S_q(r)}{[S_q(r)]_{max}}$ is never going to be exactly $1$. We
measure the deviation from homogeneity using the measure
$R_q(r)=1-\frac{S_q(r)}{[S_q(r)]_{max}}$ and consider a distribution
to be homogeneous when this deviation for different order $q$ are
statistically indistinguishable. We set an arbitrary but sufficiently
small limit ($\sim 10^{-3}$) for the observed differences below which
we treat them to be identical. We consider $q$ values upto $10$
keeping in mind the finite and discrete nature of the
distributions. However it should be also noted that some fluctuations
in the number count would be present in any discrete and finite
distributions. Even a homogeneous random distribution is expected to
exhibit some degree of inhomogeneity on small scales. The effect of
Poisson noise would gradually diminish with increasing length scales
due to the increase in the number counts. We expect this Poisson noise
to be also present in the analysis of the galaxy
distributions. Keeping this in mind, we also calculate the Renyi
divergence between the SDSS galaxy distribution and corresponding mock
random distributions on different length scales.

The Renyi divergence \citep{renyi61} of order $q$ is defined as,
\begin{eqnarray}
D_{q}(r) & = & \frac{1}{q-1} \log \sum_{i=1}^{M(r)} \, \frac{f^q_{i,r}}{g^{q-1}_{i,r}}
\label{eq:six}
\end{eqnarray}\\
where $g_{i,r}$ is the probability of finding a randomly chosen point
in the mock random sample within a radius $r$ from the location of the
$i^{th}$ galaxy in the real galaxy distribution. The Renyi divergence
for the value $q=1$ can be obtained by taking a limit $q \rightarrow
1$ which gives the Kullback-Liebler divergence \citep{kullback61}.

  The Renyi divergence can provide an alternative definition of the
  scale of homogeneity. It compares two distributions and would be
  ideally zero when the two distribution are identical. The galaxy
  distributions are certainly not identical with the homogeneous
  Poisson distribution. However if galaxy distribution becomes
  homogeneous on larger length scales then any small signal of
  inhomogenety present on and above this length scales may be purely
  an outcome of the Poisson noise. The Poisson distributions are
  homogeneous by definition and the inhomogeneity at any length scale
  in these distributions arises due to the Poisson noise alone. When
  compared with a homogeneous random distribution, a very small value
  of the Renyi divergence on a given length scale suggests that the
  galaxy distribution is also homogeneous on that length scale. \\

  It may be also noted that the different outcomes in \autoref{eq:two}
  are expected to be independent of each other. However this
  assumption does not hold due to the overlaps between different
  spheres. This becomes particularly important on larger length scales
  where a greater degree of overlap between the spheres may
  significantly affect the measurement of the scale of homogeneity in
  the present framework.  We also analyze our data using independent
  volumes in order to address this issue. We superpose the galaxy
  sample with an uniform three dimensional rectangular grid and
  identify all the grid cells that lie completely inside the boundary
  of the sample. These grid cells or voxels are independent as there
  are no overlaps between them. The galaxy counts in these voxels are
  used to calculate the Renyi entropy and Renyi divergence. The Renyi
  entropy and Renyi divergence can be calculated following equations
  similar to the \autoref{eq:five} and \autoref{eq:six}. The sum in
  these equations are now carried out over all the available voxels
  ($N(d)$) within the sample. One can vary the grid size $d$ and
  calculate the Renyi entropies and Renyi divergences associated with
  the distribution on different length scales. Albeit the number of
  the available voxels would decrease significantly on larger length
  scales. A somewhat higher tolerance values for $R_q(d)$ and $D_q(d)$
  should be allowed to identify the scale of homogeneity in this
  framework.

\section{Data}

We apply the method to data from the SDSS and Monte-Carlo simulations
of the homogeneous Poisson point process.

\subsection{SDSS Data}

\begin{figure*}[htbp!]
\centering
\includegraphics[width=7cm]{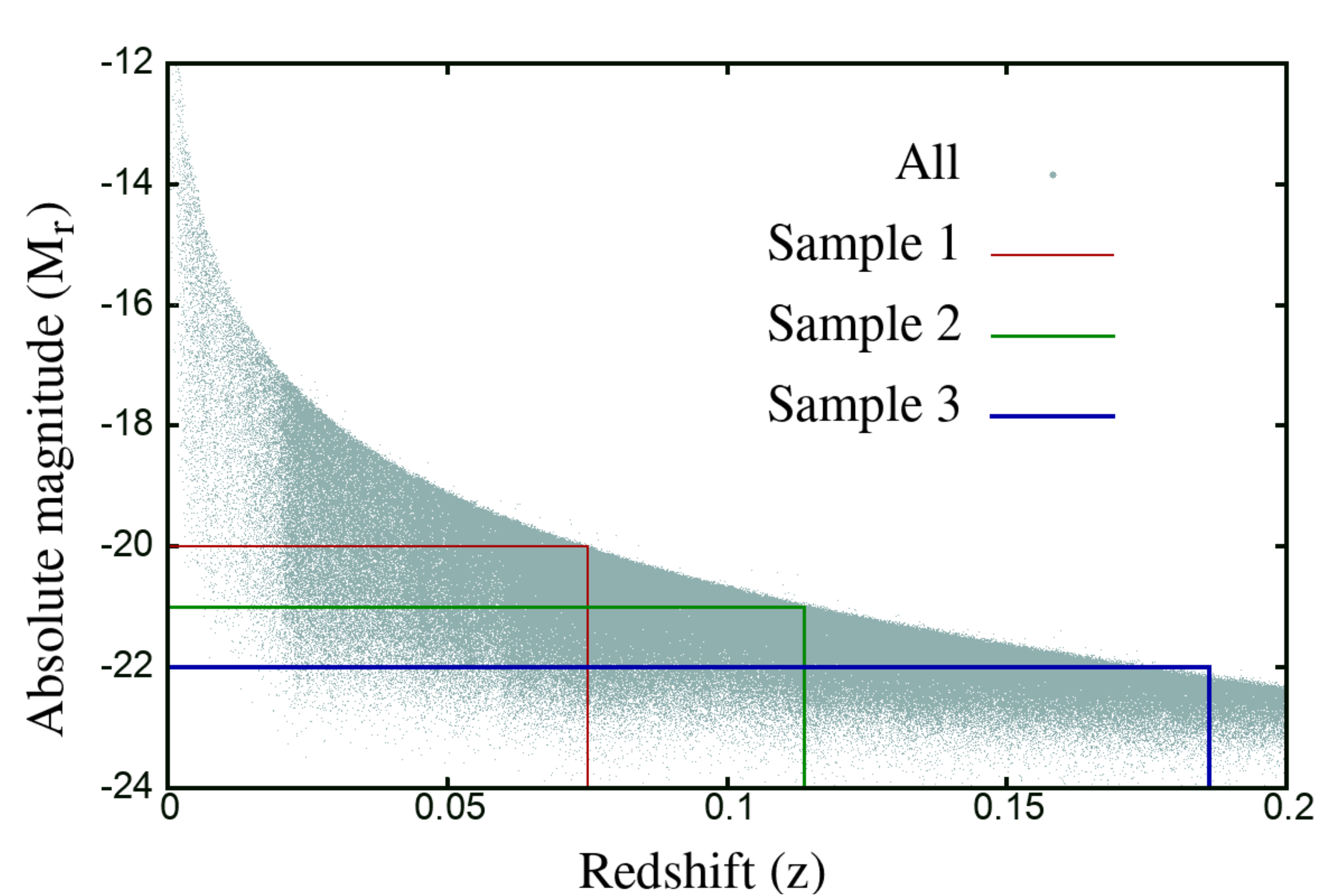}\hspace{0.25cm} 
\vspace{0.25cm}
\includegraphics[width=7cm]{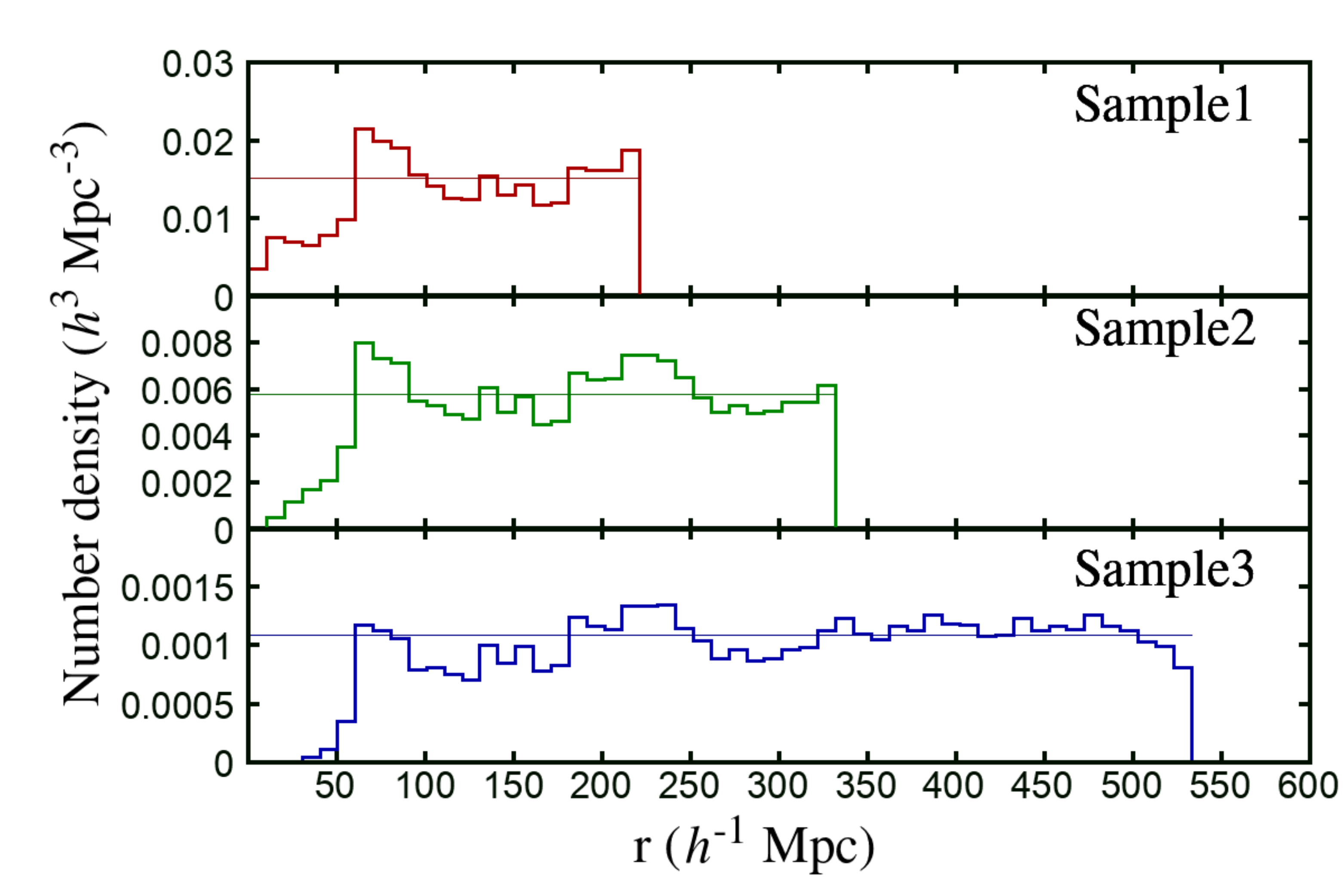}
\caption{The left panel shows the definition of the three
  volume limited samples in the absolute magnitude-redshift plane. The
  right panel shows the variation of the comoving number density
  with radial distance in the three samples. The number densities are
  calculated in shells of uniform thickness $10 \hmpc$. The horizontal
  lines in the right panel show the mean density in the
  respective samples.}
\label{fig:samples}
\end{figure*}

The Sloan Digital Sky survey (SDSS) \citep{york} is the most
successful redshift survey in the history of astronomy. The SDSS
photometric camera is described in Gunn et al \citep{gunn98}. Strauss
et al. \citep{strauss02} describe the target selection algorithm for
the SDSS main sample. The SDSS provides spectral sky coverage of
$9,376$ square degrees targeting around 3 million galaxies in its
third phase. We extract data from the sixteenth data release (DR16)
\citep{ahumada} of SDSS using a SQL query in the
CasJobs \footnote{https://skyserver.sdss.org/casjobs/}. We identify a
contiguous region in the northern galactic hemisphere between the
right ascension $125^\circ$ and $235^\circ$ and declination $0^\circ$
and $60^\circ$. We prepare three different volume limited samples with
{\it r}-band Petrosian absolute magnitude limit $M_r \leq -20$, $M_r
\leq -21$ and $M_r \leq -22$. We name these samples as Sample 1,
Sample 2 and Sample 3 respectively. The three volume limited samples
are defined in the redshift-absolute magnitude plane in the left panel
of \autoref{fig:samples}. The right panel of \autoref{fig:samples}
shows the comoving number density in these samples as a function of
radial distance. The properties of these volume limited samples are
summarized in \autoref{tab:samples}.

For conversion of redshift to comoving distance we use the
$\Lambda$CDM cosmological model with $\Omega_{m0}=0.315$,
$\Omega_{\Lambda0}=0.685$ and $h=0.674$ \citep{planck18}.

\subsection{Random data}
We simulate a set of mock random distributions using Monte Carlo
simulations described in Pandey \citep{pandey21}. We generate 10 mock
random realizations for each volume limited sample from the
homogeneous Poisson point processes. The mock random data sets contain
same number of points as there are galaxies in the volume limited
sample. The geometry and volume of each random data sets are also
exactly identical with the actual SDSS samples.

\begin{figure*}[htbp!]
\centering
\includegraphics[width=7cm]{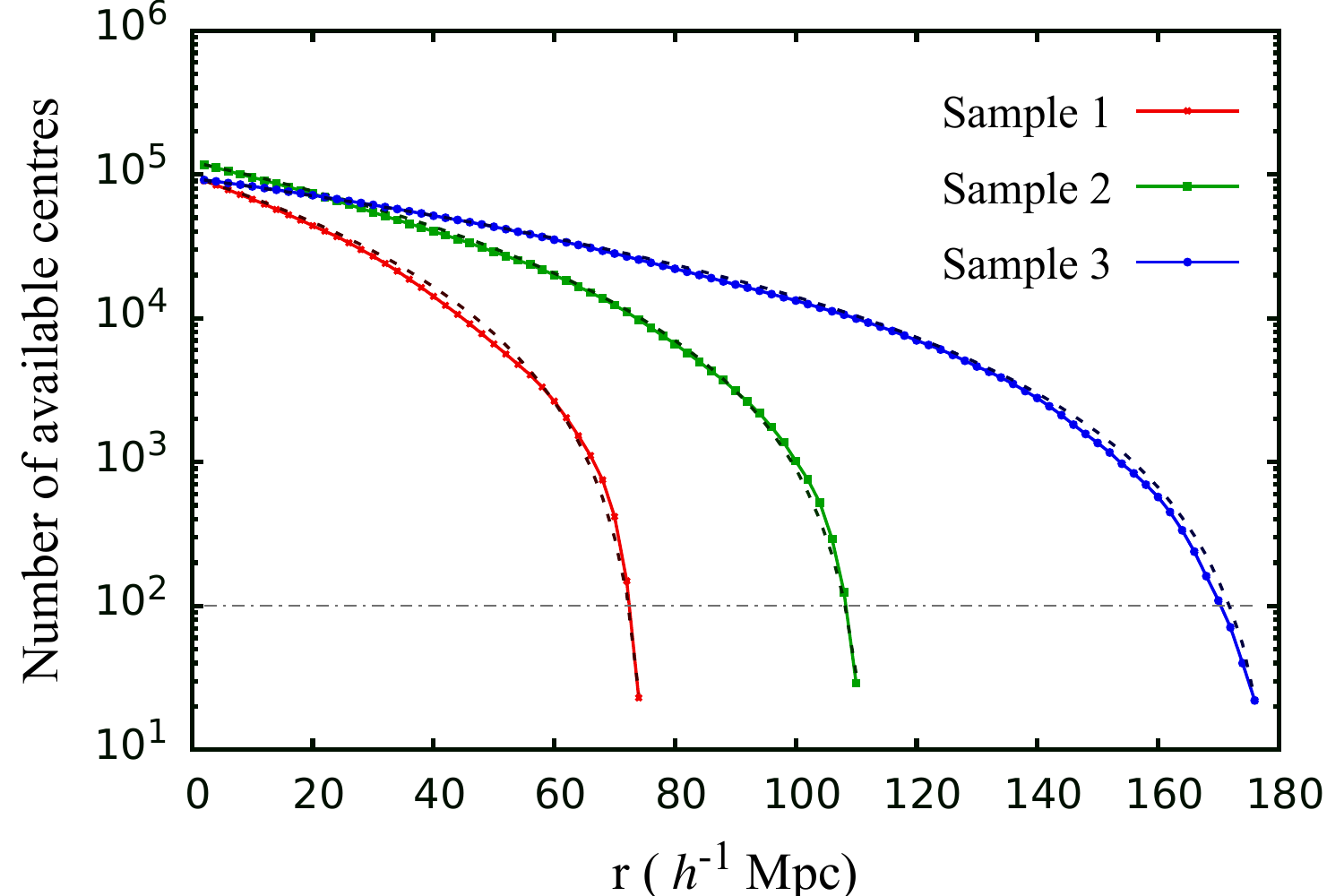}\hspace{1cm} 
\vspace{0.4cm}
\includegraphics[width=7cm]{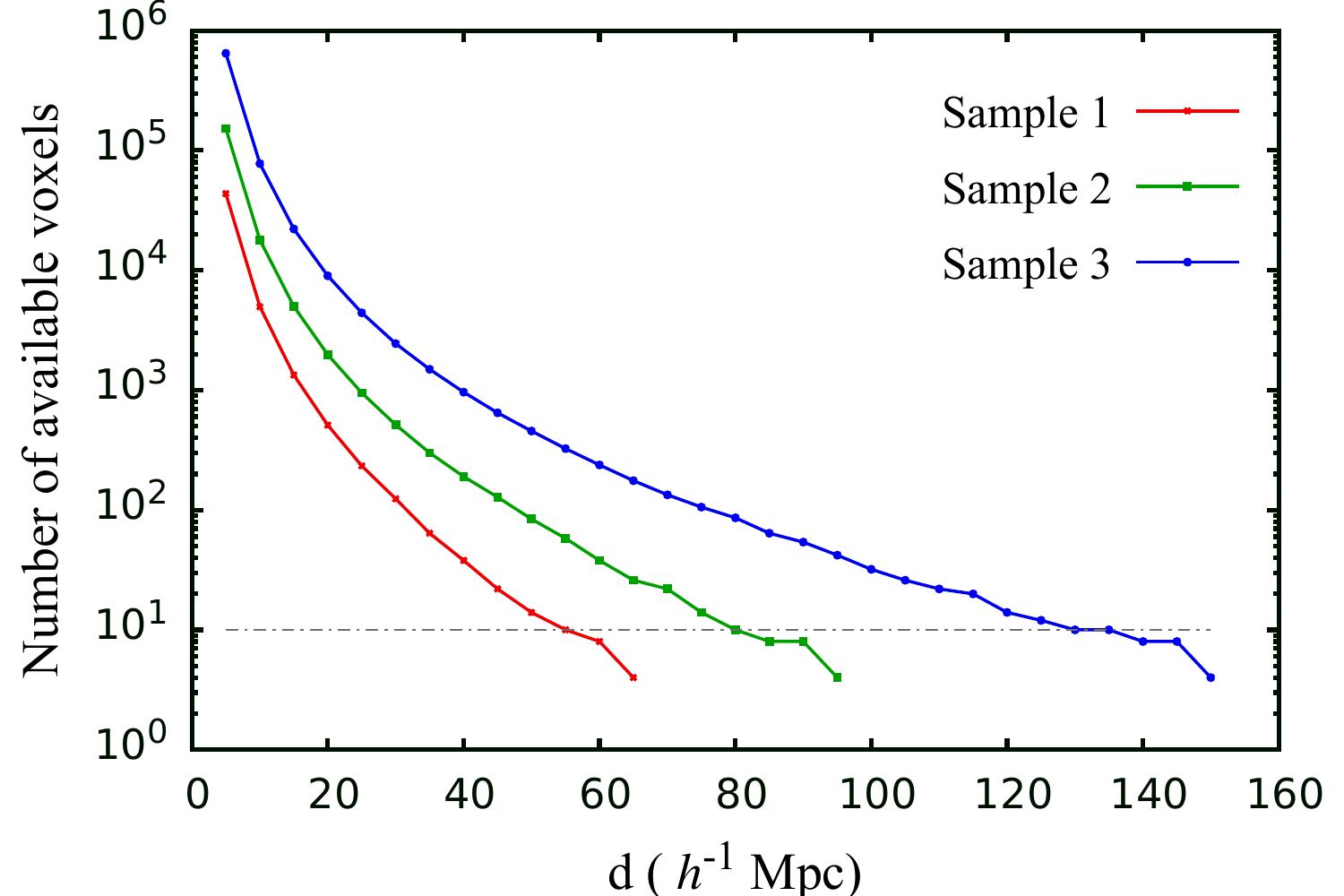}
\caption{The left panel of this figure shows the number of centres
  available on different length scales for Sample 1, Sample 2 and
  Sample 3 of the SDSS data. The solid lines and the dashed lines in
  this panel show the results for the real galaxy samples and the mock
  random samples respectively. The right panel shows the number of
  independent voxels available at each length scale for these galaxy
  samples.}
\label{fig:centres}
\end{figure*}

\begin{figure*}[htbp!]
\centering
\includegraphics[width=7cm]{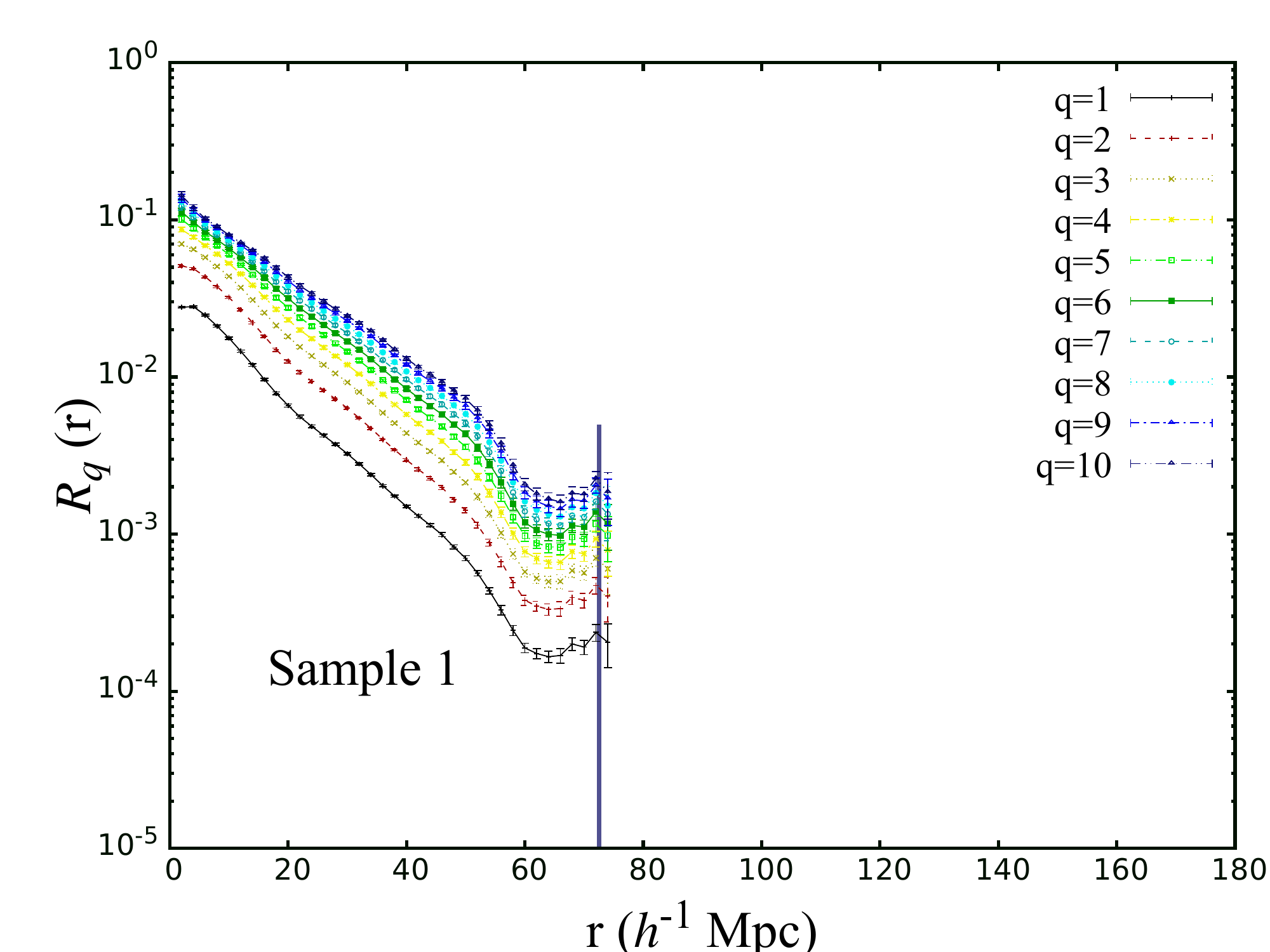}\hspace{0.25cm} 
\vspace{0.25cm}
\includegraphics[width=7cm]{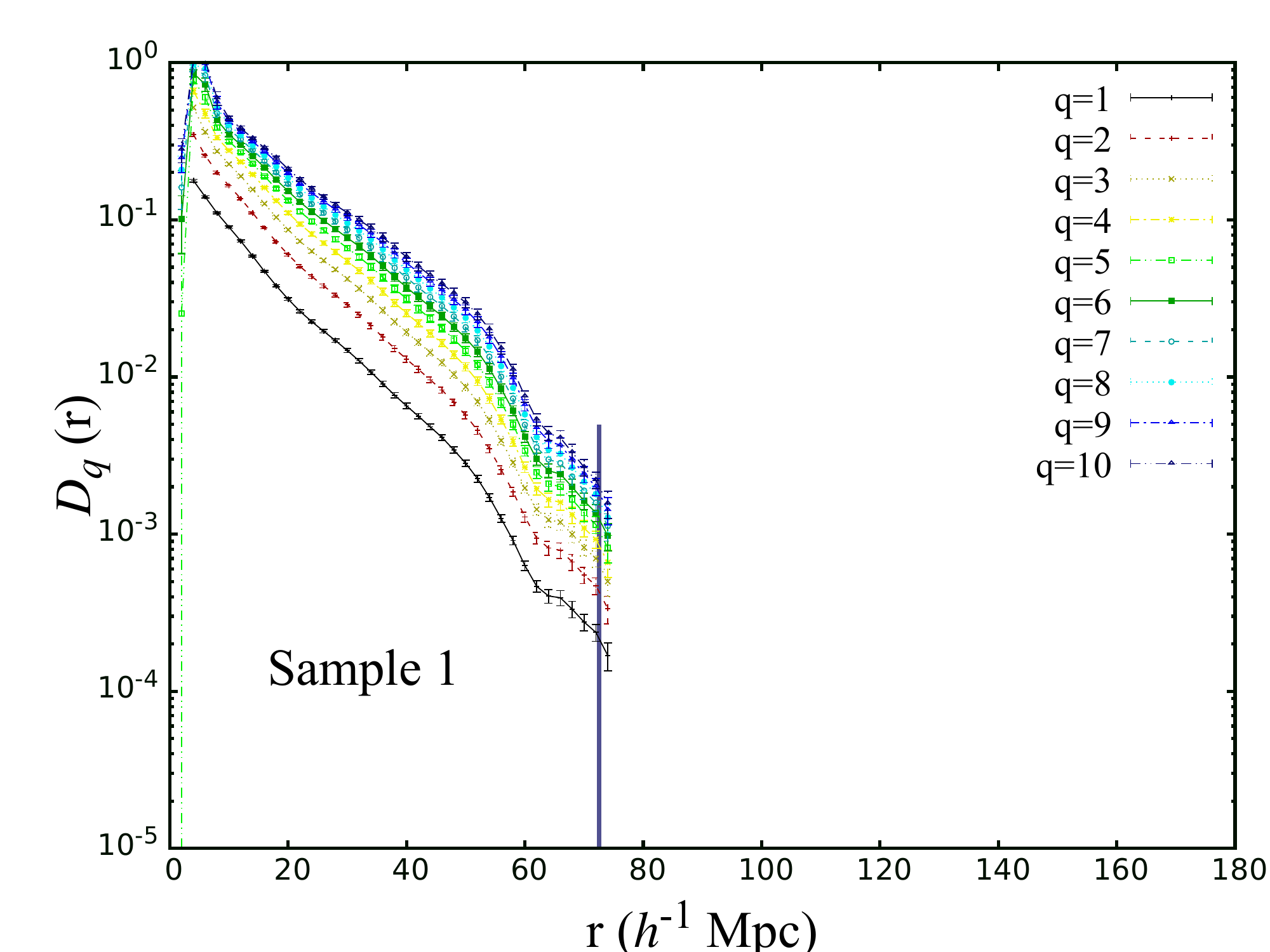}
\includegraphics[width=7cm]{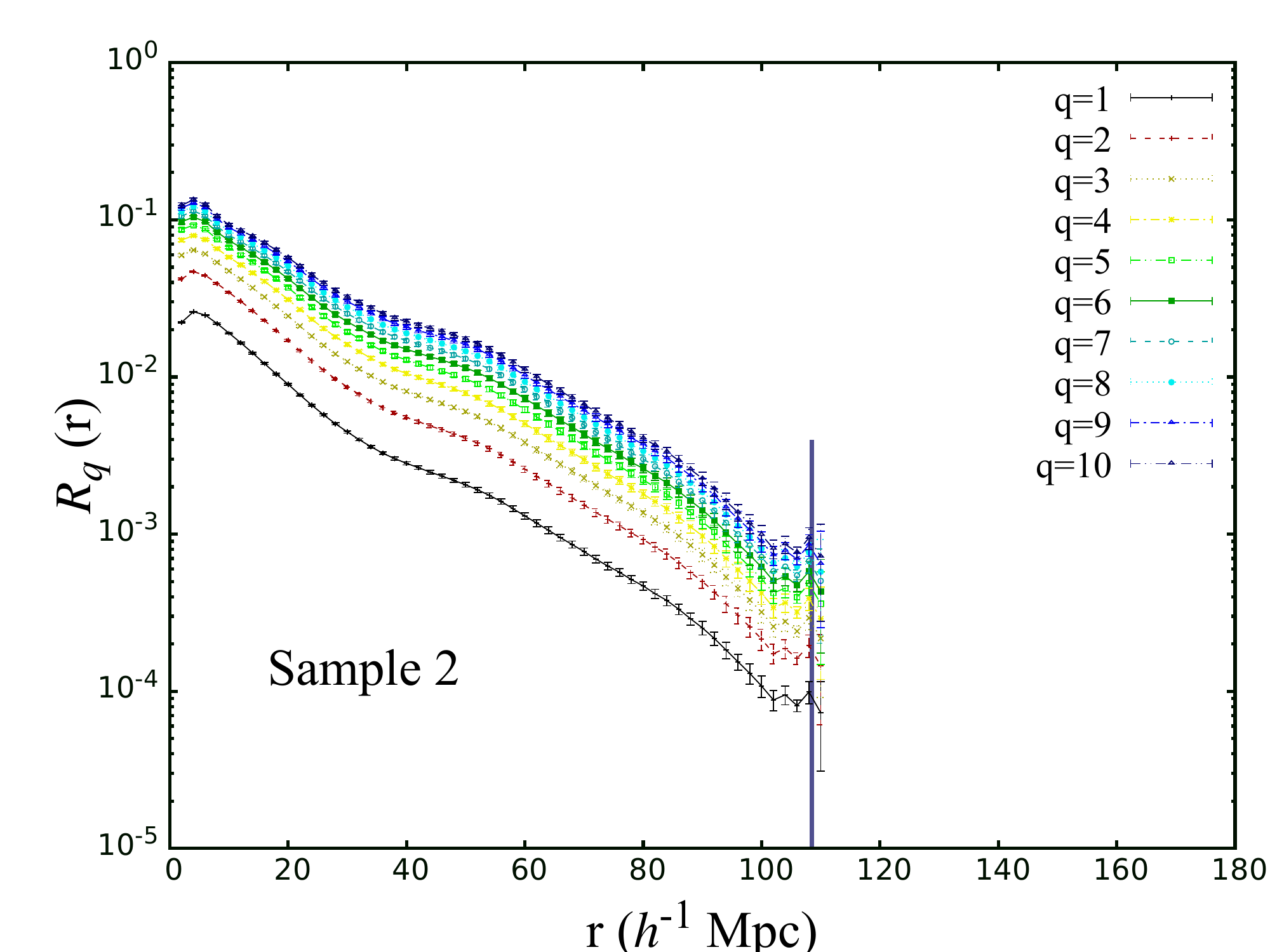}\hspace{0.25cm}
\includegraphics[width=7cm]{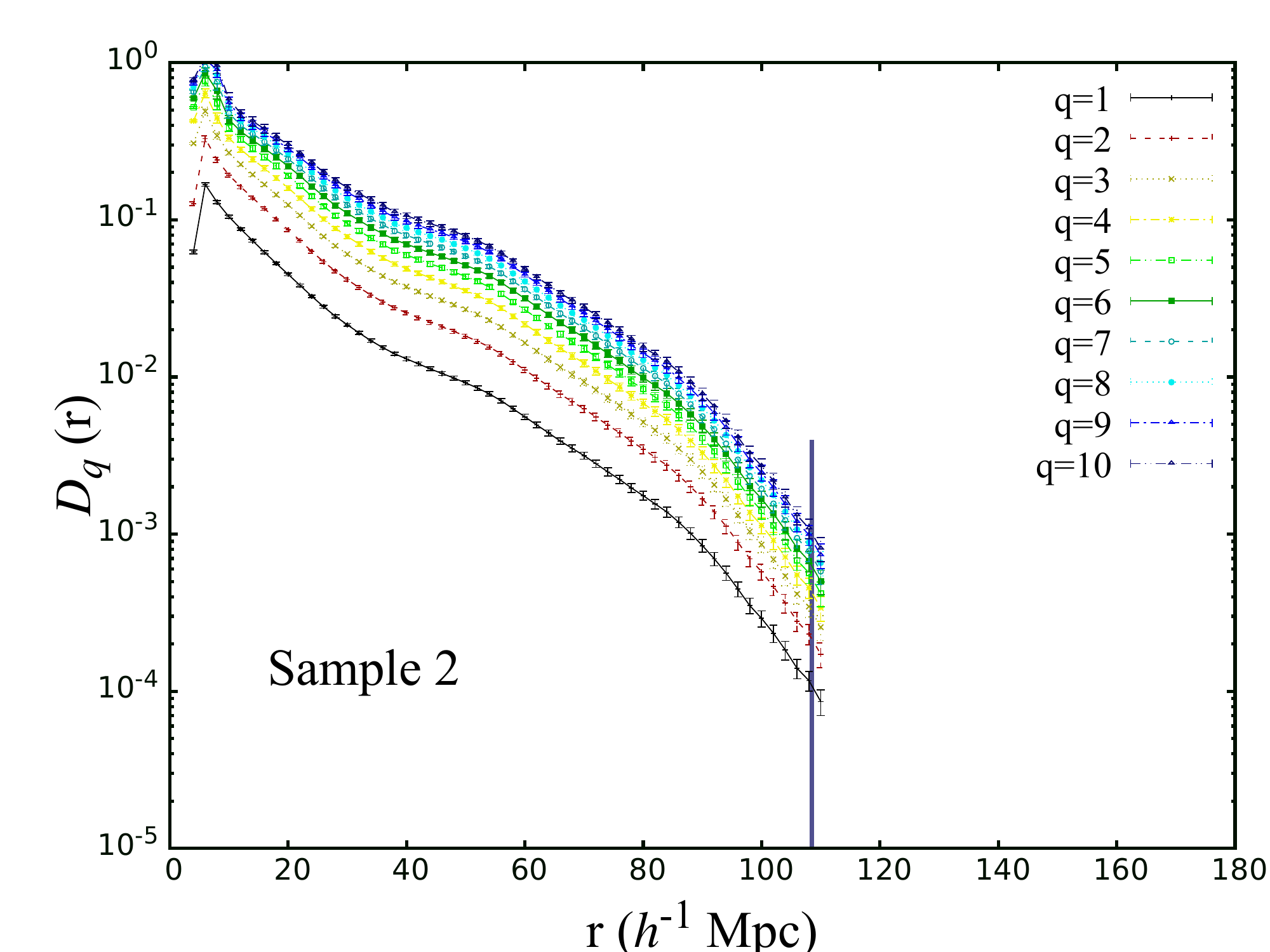}
\includegraphics[width=7cm]{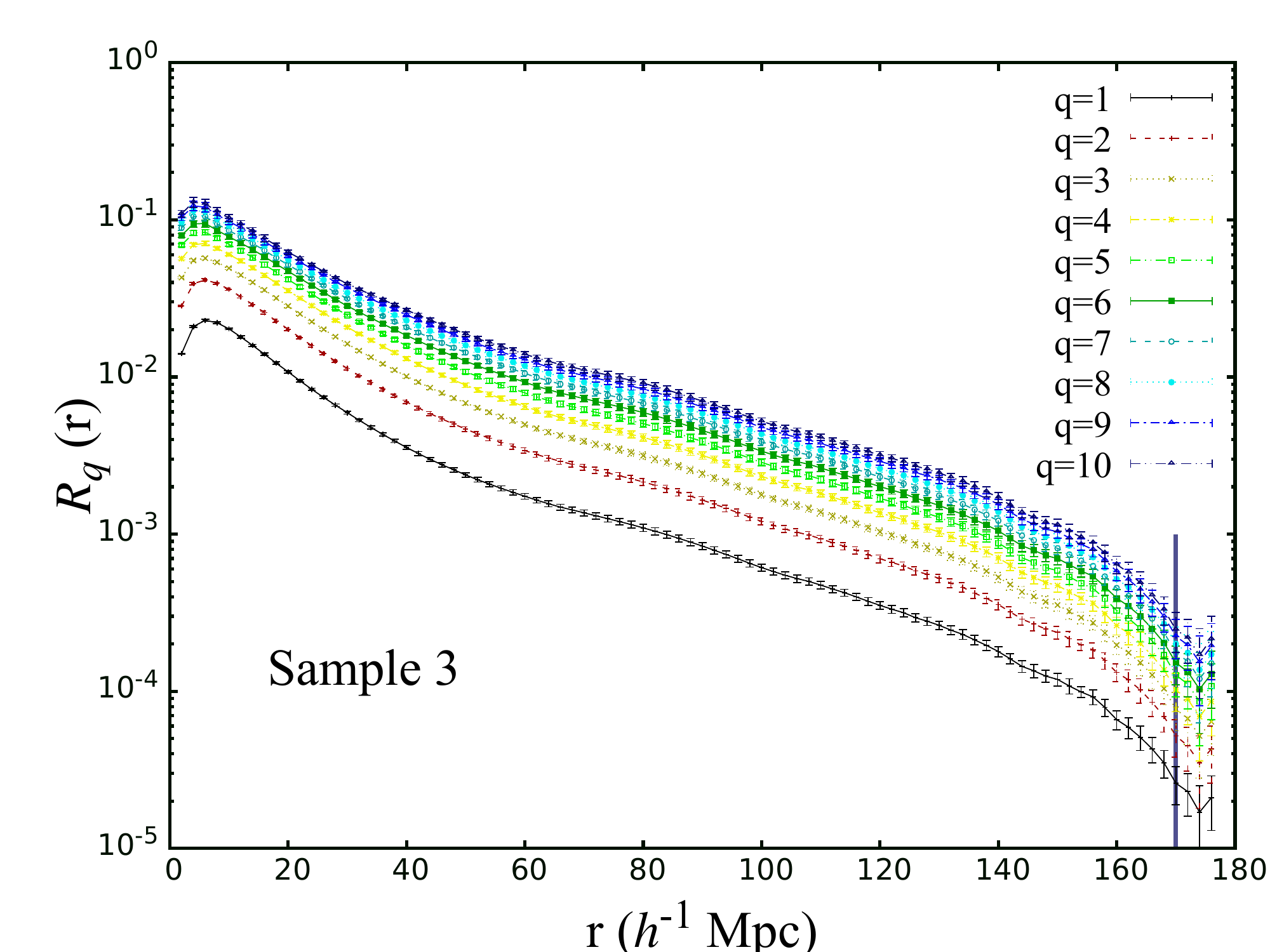}\hspace{0.25cm}
\includegraphics[width=7cm]{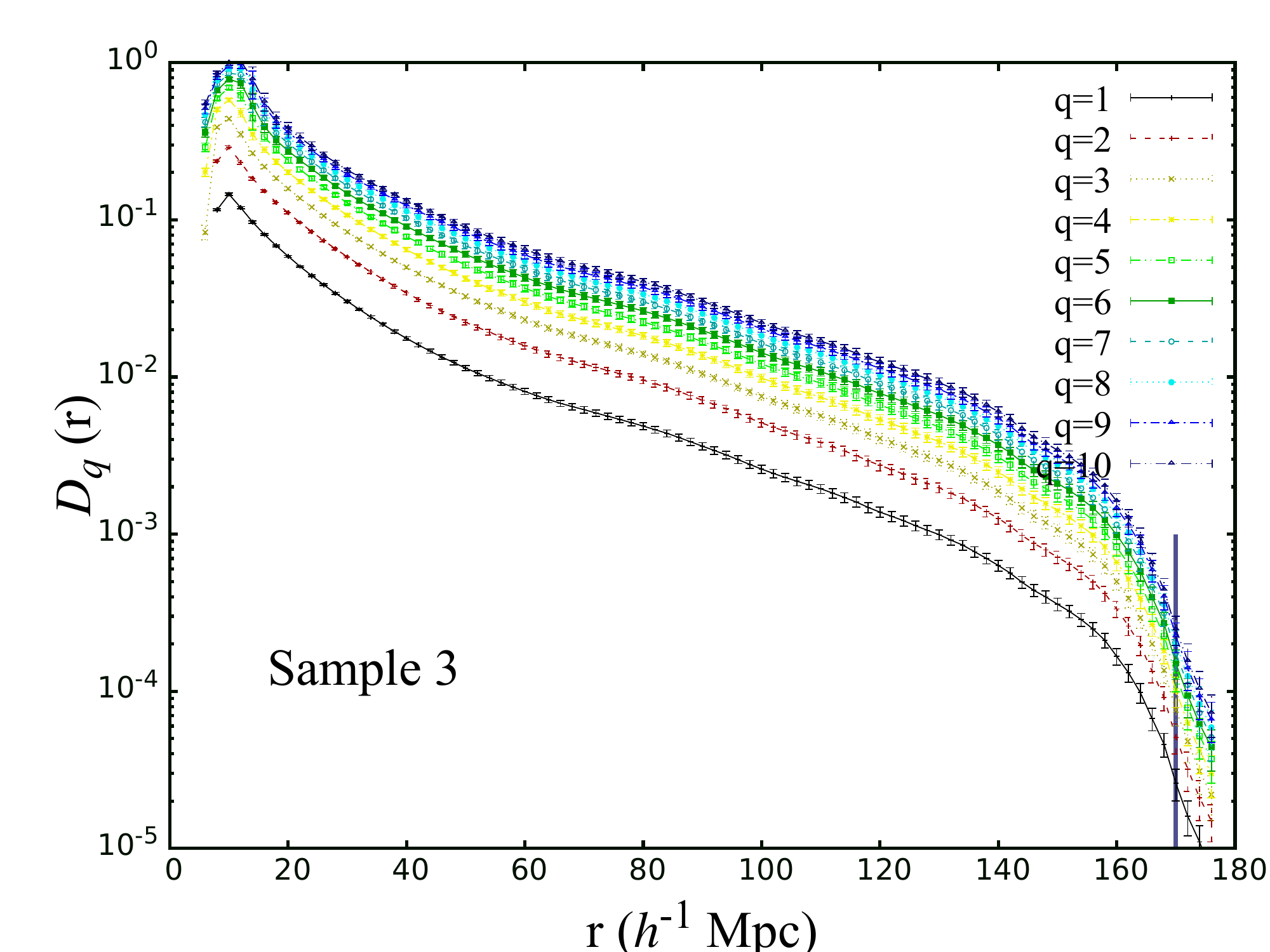}
\caption{The top left, middle left and bottom left panels show
  $R_q(r)$ as a function of length scales for different values of $q$
  in Sample 1, Sample 2 and Sample 3 respectively when the number
  counts are measured using overlapping spheres. The top right, middle
  right and bottom right panels show the Renyi divergence $D_q (r)$ as
  a function of length scale for these samples. In each case, the
  $1-\sigma$ errorbars for $R_q(r)$ are estimated using $10$ jackknife
  sub-samples. $10$ mock Poisson samples are used to estimate the
  $1-\sigma$ errorbars for $D_q(r)$ in each galaxy sample. The
  vertical solid line on the x-axis of each panel corresponds to the
  length scale at which $100$ overlapping spheres are available in the
  respective sample.}
\label{fig:renyi}
\end{figure*}

\begin{figure*}[htbp!]
\centering
\includegraphics[width=7cm]{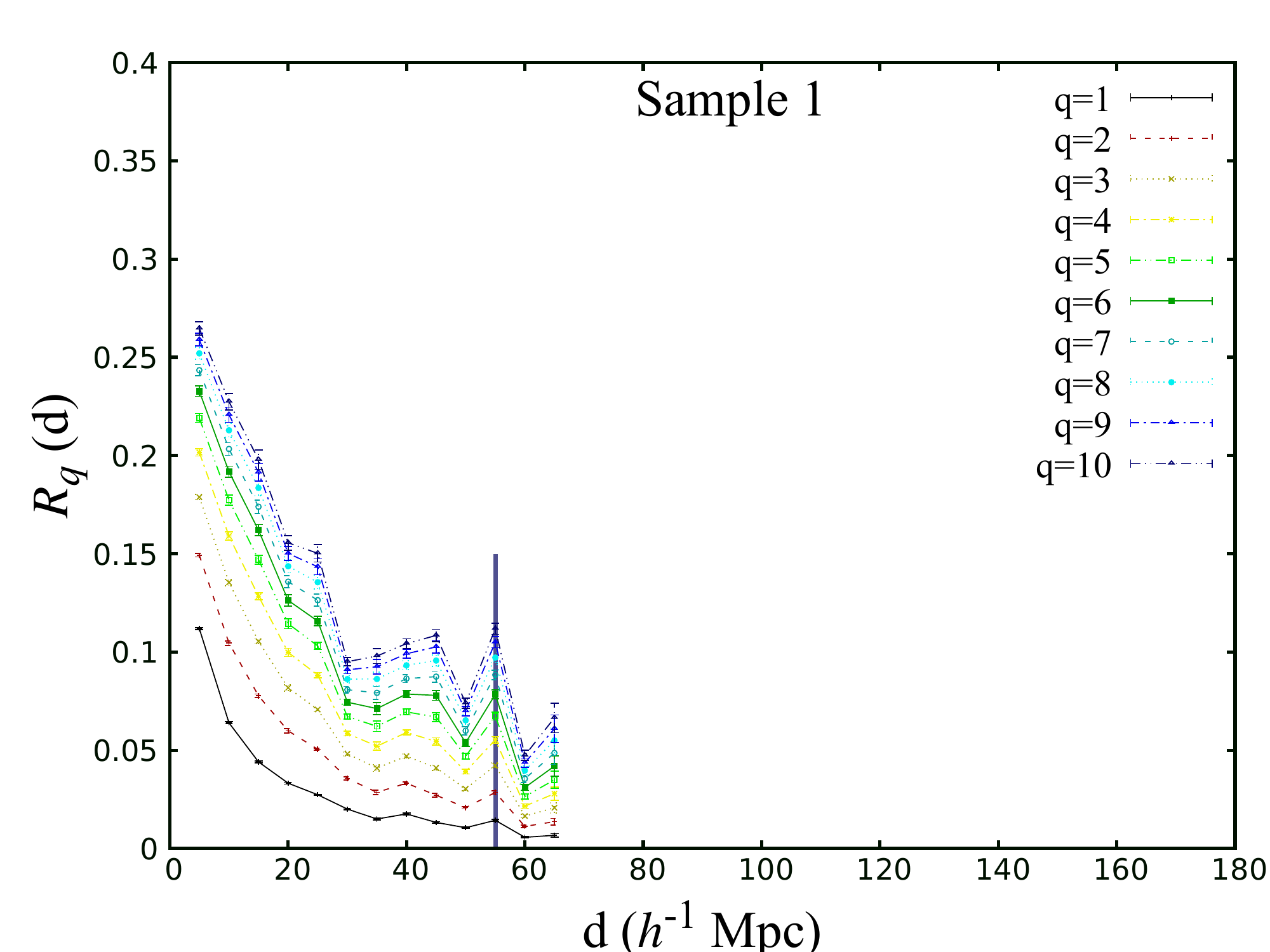}\hspace{0.25cm} 
\vspace{0.25cm}
\includegraphics[width=7cm]{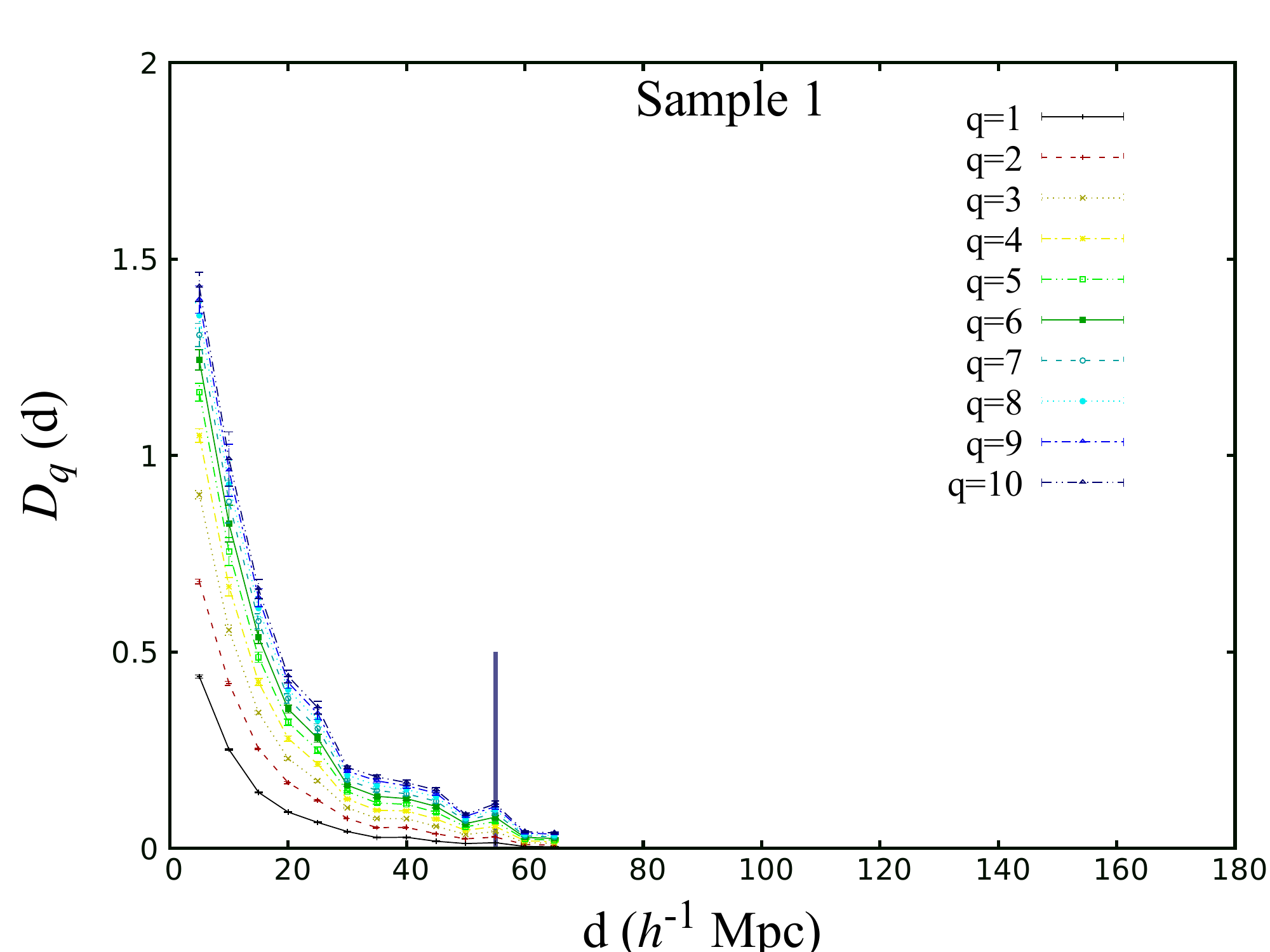}
\includegraphics[width=7cm]{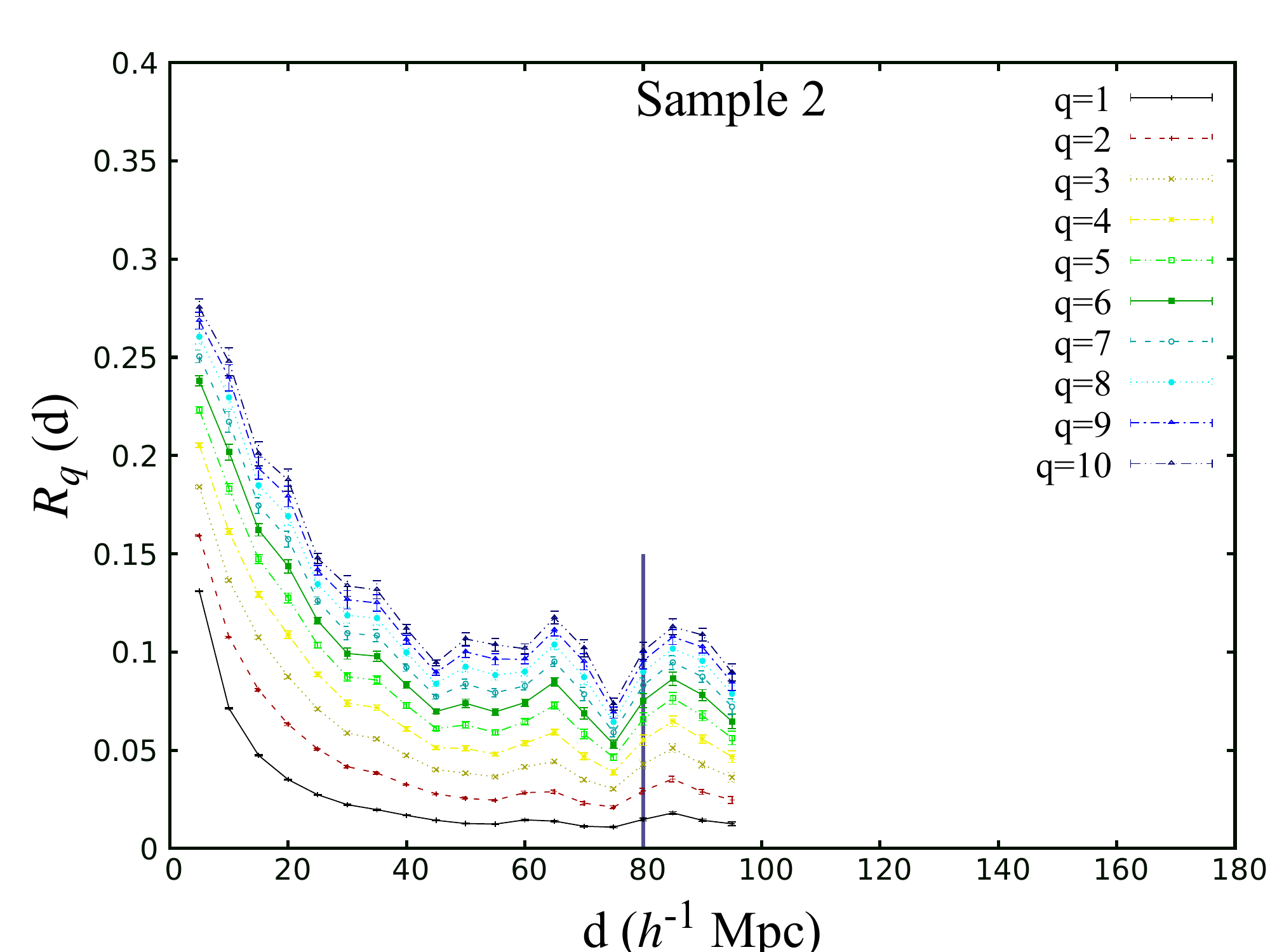}\hspace{0.25cm}
\includegraphics[width=7cm]{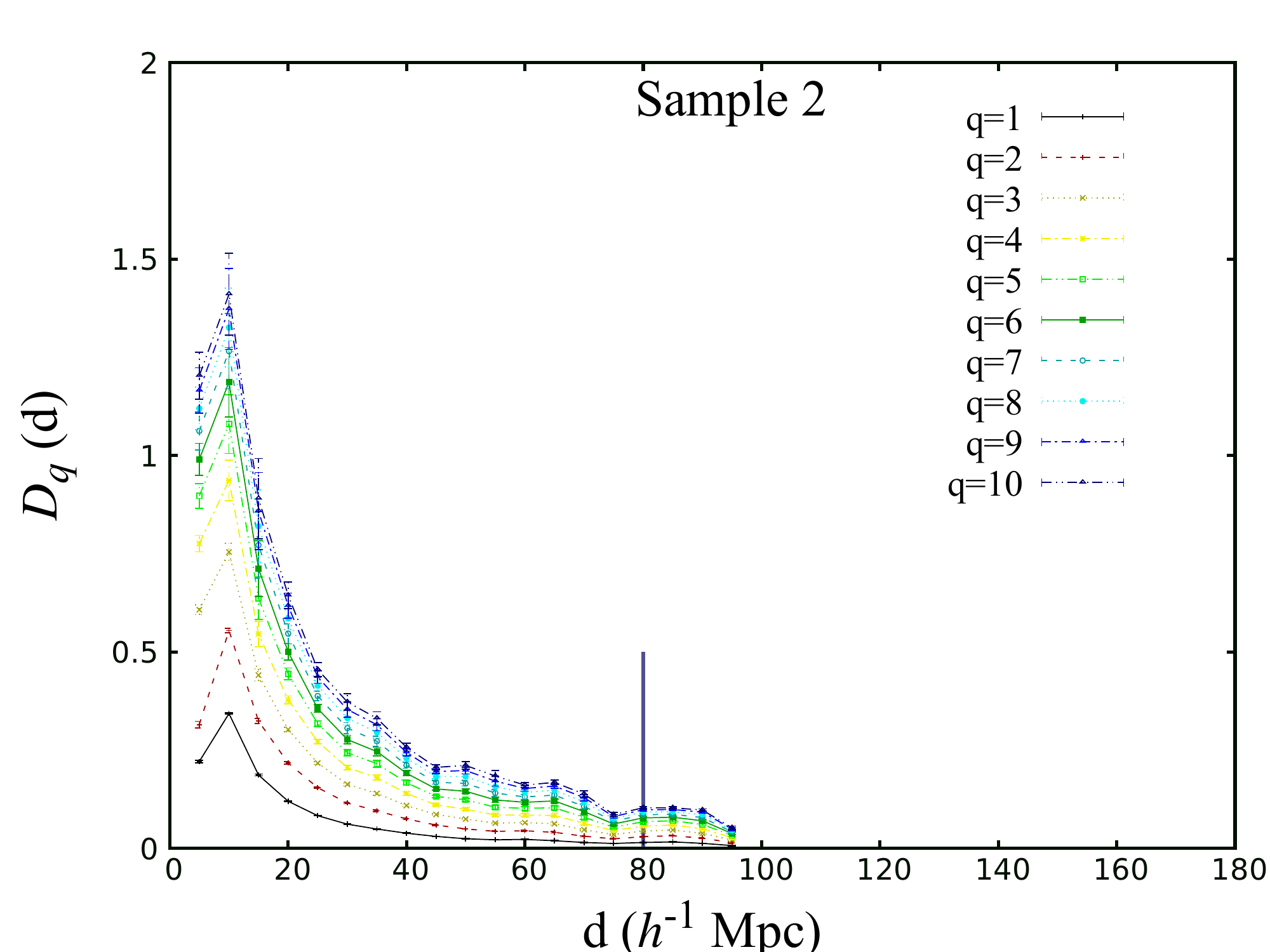}
\includegraphics[width=7cm]{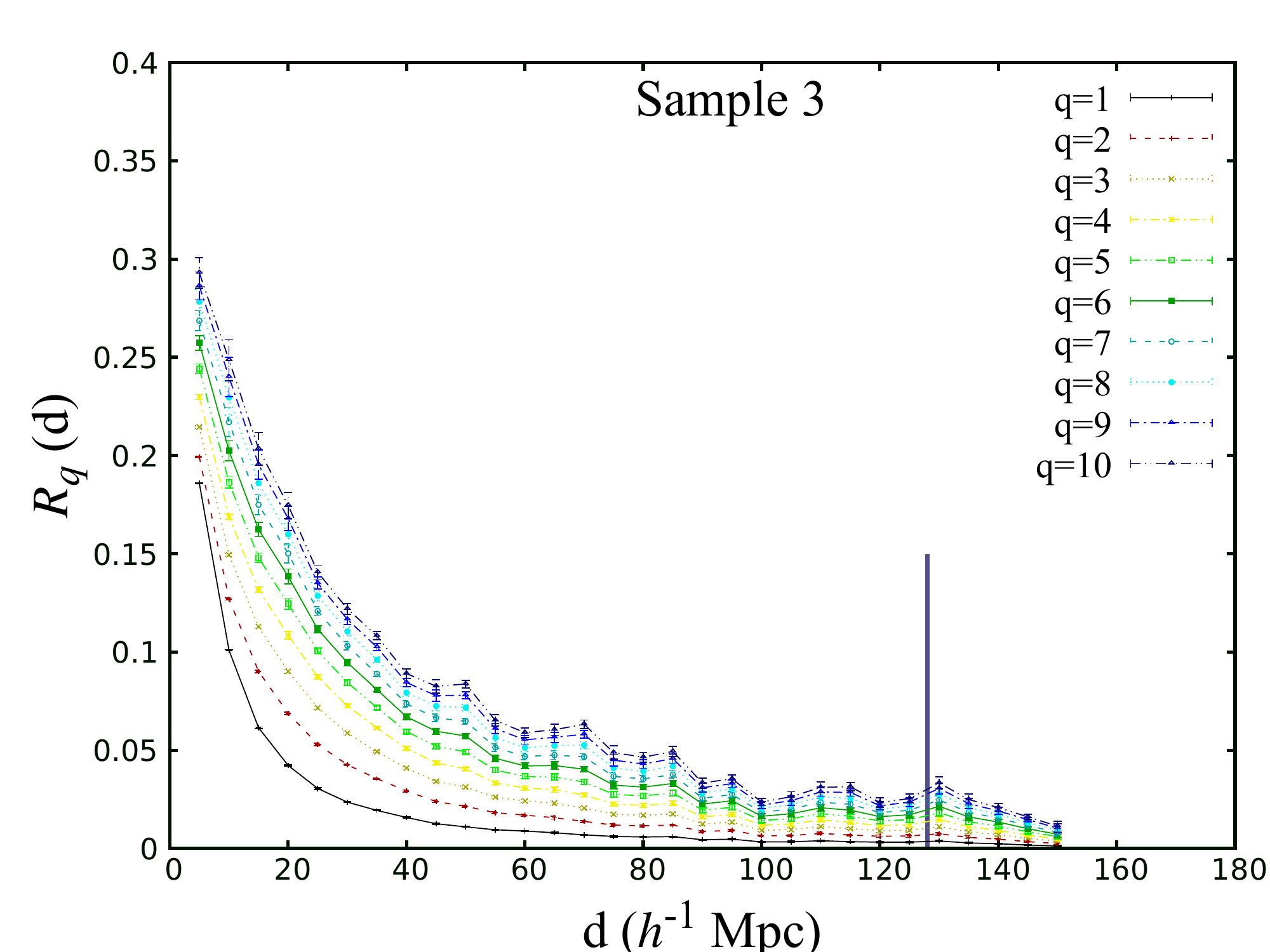}\hspace{0.25cm}
\includegraphics[width=7cm]{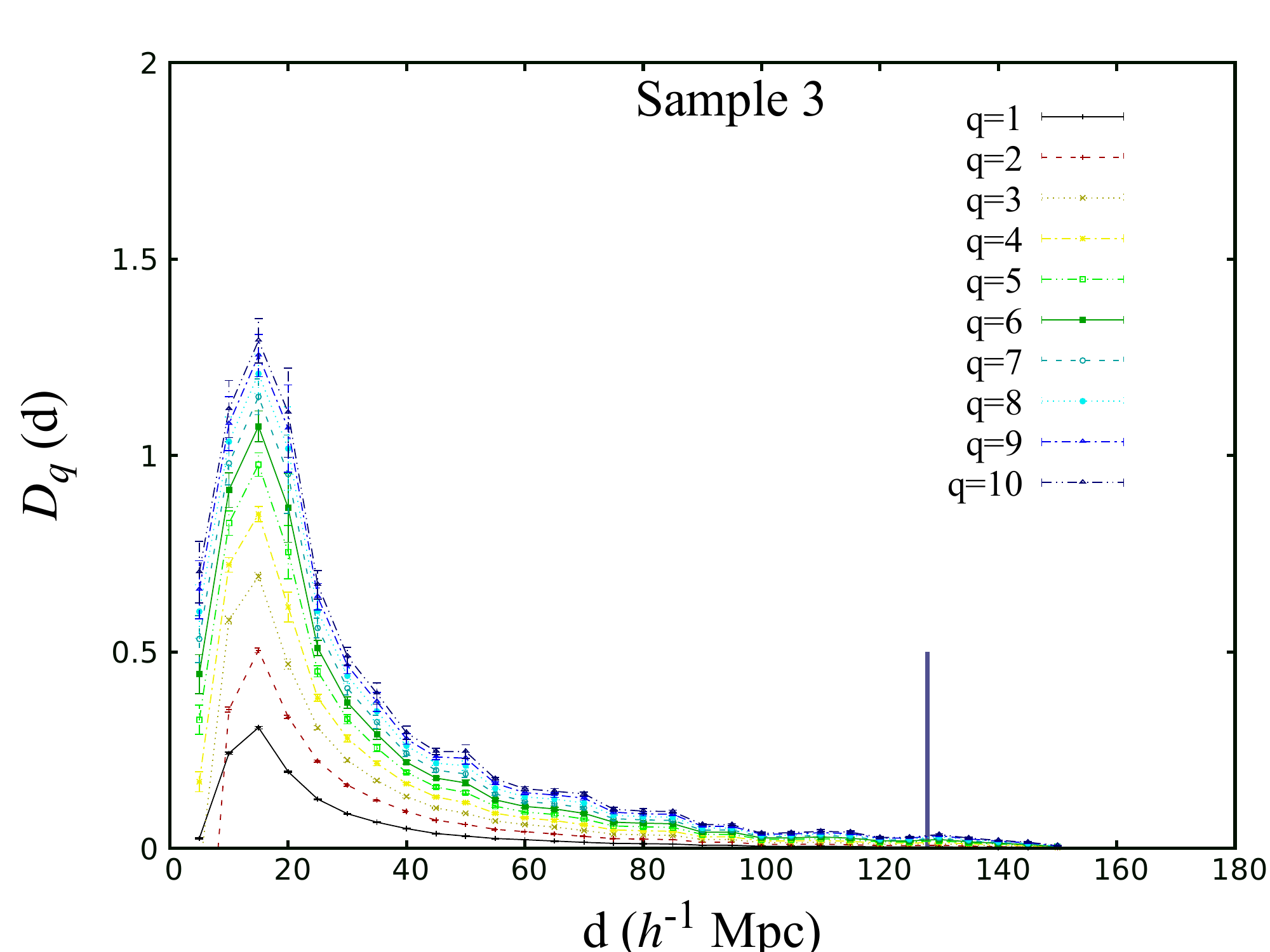}
\caption{Same as \autoref{fig:renyi} but when the number counts are
  measured in independent voxels. The vertical solid line on the
  x-axis of each panel corresponds to the length scale at which $10$
  independent voxels are available in the respective sample.}
\label{fig:renyi_ind}
\end{figure*}

\begin{table*}{}
\caption{This table summarizes the properties of our volume limited samples
  from SDSS.}
\label{tab:samples}
\begin{tabular}{ccccccc}
\hline
	& Absolute			& Redshift 	& Number of 	& Volume of 	& Number	& Mean \\
	& magnitude range 	&range 		&Galaxies 		&the region  	& density 	& separation\\
	&	&	&(N)			&$(h^{-1}\, {\rm Mpc})^{3}$		&$(h^{-1}\, {\rm Mpc})^{-3}$	&$(h^{-1}\, {\rm Mpc})$\\
\hline
Sample 1	&$ M_{r} \leq -20$		& $ z \leq 0.075$	&$90618$	&$5.96\times10^6$	&$1.52\times10^{-2}$	&$4.03$	\\
&&&&&&\\
Sample 2	&$ M_{r} \leq -21$		& $ z \leq 0.114$	&$117058$	&$2.02\times10^7$	&$5.79\times10^{-3}$	&$5.57$	\\
&&&&&&\\
Sample 3	&$ M_{r} \leq -22$		& $ z \leq 0.186$	&$91626$	&$8.87\times10^7$	&$1.09\times10^{-3}$	&$9.70$	\\
\hline
\end{tabular}
\end{table*}

\section{Results}
\subsection{Analysis with the overlapping spheres}
We show the number of available centres as a function of length scales
in the three SDSS volume limited samples in the left panel of
\autoref{fig:centres}. The left panel of \autoref{fig:centres} shows
that the number of centres in Sample 1 decreases rapidly with
increasing length scales due to its smaller size. The number of
centers drops to $\sim 100$ at $\sim 70 \hmpc$ in the Sample 1. The
number of centers falls more slowly in Sample 2 reaching $100$ at
around $110 \hmpc$. The decrease is much slower in Sample 3 due to its
larger volume. Sample 3 has an order of magnitude larger volume than
Sample 1 (\autoref{tab:samples}). The number of centers reach to $100$
at $\sim 170 \hmpc$ in this sample. The number of valid centres for
the corresponding mock random datasets are shown together using dashed
lines in the same panel. Their similarity suggests that the
availability of the valid centers on different length scales are
largely decided by the geometry and volume of the sample and the
number of galaxies contained within it. In principle, it should also
depend on the nature of the clustering in the data but the dependence
is much weaker \citep{pandey13}. We show the availability of the
number of independent voxels as a function of length scale for the
three samples in the right panel of the same figure. The number of
voxels for the actual data and random data are exactly same as it
depends only on the geometry and volume of the sample.

We now measure the Renyi entropies of different order for the SDSS and
random data. We show $R_q(r)$ as a function of length scales $r$ in
\autoref{fig:renyi}. The top left, middle left and bottom left panels
of this figure show the results for Sample 1, Sample 2 and Sample 3
respectively. The corresponding Renyi divergences are shown in the top
right, middle right and bottom right panels of the same figure.

The top left panel of \autoref{fig:renyi} shows that the deviation
from homogeneity as quantified by $R_q(r)$ decreases with increasing
length scales for each $q$ values. The higher $q$ values show a larger
deviation at smaller length scales. The differences diminish with the
increasing length scales and $R_q(r)$ for different $q$ values becomes
of the order of $\sim 10^{-3}$ at $60 \hmpc$. This may apparently
indicate a transition to homogeneity at $\sim 60 \hmpc$. However, this
can be misleading. The $R_q(r)$ curve for each $q$ value for Sample 1
shows a sudden change in its slope around $50 \hmpc$. This is caused
by a progressive overlap of the measuring spheres eventually enforcing
an artificial homogeneity. A solid vertical line is used to indicate
the length scale at which only $100$ measuring spheres are available
for the analysis. The number $100$ is arbitrarily chosen keeping in
mind the severity of the contamination with a very a small number of
spheres.  Though this occurs at $\sim 70 \hmpc$ for Sample 1, a sudden
change in the slopes of the $R_q(r)$ curves are noticeable even before
this length scale. This indicates that the process is gradual and
becomes dominant on larger length scales.

The Renyi divergences for Sample 1 are shown in the top right panel of
this figure. The Renyi divergence $D_q(r)$ measures the deviation of
the galaxy distribution from a Poisson random distribution. We find
that it decreases with the increasing length scales and $D_q(r)$ for
different $q$ values become $\sim 10^{-2}$ or smaller on a length
scale of $60 \hmpc$. This amounts to $\sim 100$ time reduction in the
peak value of the Renyi divergence for each $q$. However the
predominance of the overlapping bias on such length scales in Sample 1
does not allow us to identify this length scale as the scale of
homogeneity. The fact that we see a deviation from homogeneity upto
$60 \hmpc$ in the SDSS Sample 1 indicates the presence of real
inhomogeneity on such length scales in the galaxy distribution. The
apparent transition to homogeneity at $60 \hmpc$ for the Sample 1 is
forced by the confinement and overlapping bias
\citep{pandey13,kraljic}. This is clear from the \autoref{fig:centres}
which shows that the number of available centres in Sample 1 shows a
sudden drop beyond $60 \hmpc$. Pandey \citep{pandey13} pointed out
that the available valid centers in any finite dataset would gradually
migrate towards the centre of the volume with increasing length
scales. This would lead to a larger degree of overlap between the
measuring spheres undermining the actual degree of inhomogeneities
present on such length scales. Thus the small number of spheres (a few
hundred to less than hundred) available beyond $60 \hmpc$ in Sample 1
may have a large degree of overlap. These spheres are probing nearly
the same volume thereby reflecting very little differences between
them. Such artificial homogeneity may be commonplace in all the number
count based methods applied to the finite datasets.

In the middle left and middle right panels of \autoref{fig:renyi}, we
respectively show the $R_q(r)$ and $D_q(r)$ curves for different $q$
values in Sample 2. The peaks at certain length scales in these plots
corresponds to the mean intergalactic separation in these
distributions \citep{pandeysarkar15}. Clearly, the inhomogeneities in
SDSS Sample 2 now persists beyond $60 \hmpc$ and extends at least upto
$90-100 \hmpc$.  This is related to the fact that the inhomogeneities
can be now detected upto a larger length scales due to the greater
size of Sample 2. But despite the larger size of Sample 2, the
overlapping and confinement bias eventually overshadow the
inhomogeneities present in the galaxy distribution beyond $100 \hmpc$
\autoref{fig:centres}. The Renyi divergences for Sample 2 in the
middle right panel show that a reduction of $D_q(r)$ by a factor of
$100$ are observed on a scale of $\sim 90 \hmpc$. The inhomogeneities
may be present in the galaxy distribution beyond $90 \hmpc$.  But we
can not detect them due to the dominance of the overlapping bias
beyond this length scale. A solution to this problem is to consider
significantly larger sample of galaxies so that the overlapping bias
becomes important on much larger length scales. This would allow us to
probe the transition scale to homogeneity provided such a transition
occurs on a relatively smaller length scale.

We analyze a large galaxy sample (Sample 3, \autoref{tab:samples})
from the SDSS to explore the possibility of identifying the transition
scale to homogeneity in the observed galaxy distribution. The results
are shown in the bottom panels of \autoref{fig:renyi}. It may be noted
that the Sample 3 has a volume which is $\sim 15$ times larger than
Sample 1 and $\sim 5$ times larger than Sample 2. The overlapping bias
in Sample 3 dominates on much larger length scales at $\sim 170
\hmpc$. This enables us to probe the presence of inhomogeneties in
this sample on a length scales $<170 \hmpc$. Interestingly, we find
that the $R_q(r)$ for all the $q$ values becomes of the order of
$10^{-3}$ on a scale of $\sim 140 \hmpc$.  We note that there are a
few thousands of centres available at this radius. So this should not
be considered as a necessary consequence of the overlapping
bias. Shifting our focus to the Renyi divergences shown in the bottom
right panel, we also find that $D_q(r)$ values for different $q$ show
a reduction by a factor of $100$ on a length scale of $140\hmpc$. This
indicates that the galaxy distribution in the SDSS can be considered
as homogeneous on a length scale of $140 \hmpc$ and beyond.

\subsection{Analysis with the independent voxels}

The overlapping bias suppresses the inhomogeneities on larger length
scales. An inhomogeneous distribution may thus appear as homogeneous
under the influences of the confinement and overlapping bias. In order
to avoid these biases, we also carry out the same analysis using the
galaxy counts in independent voxels. The results are shown in
\autoref{fig:renyi_ind}. The top left anf top right panels of this
figure respectively show $R_q(d)$ and $D_q(d)$ for different $q$
values in Sample 1. Here $d$ is the size of the independent
voxels. The right panel of \autoref{fig:centres} show that the
available number of independent voxel decreases much faster (as
compared to the overlapping spheres) with the increasing length
scales. The solid vertical line on the x-axis indicates the length
scale ($\sim 55 \hmpc$) at which only $10$ independent voxels are
available within Sample 1. We do not consider the results beyond this
length scale as there are too few voxels available for the
analysis. We find that the the Renyi divergence $R_q(d)$ decreases
with the increasing length scales for each $q$ value. The results
clearly show that the $R_q(d)$ curves for different $q$ values do not
coincide with each other even on the largest length scale. This
indicates the presence of inhomogeneities upto the largest length
scale probed by Sample 1. The Renyi divergence $D_q(r)$ quantifies the
deviation from a homogeneous Poisson distribution. The $D_q(d)$ for
different $q$ values in Sample 1 decrease with the increasing length
scales. It indicates that the galaxy distribution in the SDSS tends
towards a homogeneous distribution on larger length scales. However
the difference in the $R_q(d)$ curves in the top left panel and a
noticeable difference in the $D_q(d)$ curves in the top right panel at
$55 \hmpc$ indicate the presence of inhomogeneities in the galaxy
distribution on such length scales. The middle left and middle right
panels of this figure show the results for Sample 2. Similar trends
are observed in the $R_q(d)$ and $D_q(d)$ curves for Sample 2. We can
probe upto a length scale of $80 \hmpc$ due to the larger size of
Sample 2. The results for Sample 2 suggest that smaller but
significant inhomogeneities are present even on a length scale of $80
\hmpc$. We show the results for the largest sample (Sample 3) in the
two bottom panels of \autoref{fig:renyi_ind}. The Renyi entropies
$R_q(d)$ for different $q$ values in Sample 3 are very small ($\sim
10^{-2}$) and quite close to each other on a length scale of $120
\hmpc$. The Renyi divergences $D_q(d)$ for different $q$ values in
Sample 3 reduce by a factor of nearly $100$ times (from their peak
value) at $120 \hmpc$. The different $D_q(d)$ curves suggest that the
galaxy distribution in the SDSS can be considered as quite homogeneous
on a length scale of $120 \hmpc$.

\section{Conclusions}
We analyze a set of volume limited samples from the SDSS to study the
inhomogeneities in the galaxy distribution and test the possible
existence of homogeneity on larger length scales. We measure the Renyi
entropies and the Renyi divergences as a function of length scales in
each of the galaxy distributions. The Renyi divergence is used to
quantify the deviation of the galaxy distribution from a homogeneous
Poisson distribution. We carry out our analysis with two different
schemes, one using the number counts within the overlapping spheres
and another using the number counts within the independent voxels. The
analysis of the largest galaxy sample with the overlapping spheres
indicates a homogeneity scale of $140 \hmpc$ and that with the
independent voxels suggests the scale of homogeneity to be $120
\hmpc$. Combining the results from the two analysis, we consider $140
\hmpc$ as an upper limit to the scale of homogeneity. Our analysis
thus suggests that the inhomogeneities in the galaxy distribution
decrease with the increasing length scales and a transition to
homogeneity occurs on a length scale between $120-140 \hmpc$.

We consider the Renyi entropy and the Renyi divergence upto the order
$q=10$ in this analysis. The spacing between the successive $R_q(r)$
and the $D_q(r)$ curves become narrower with the increasing values of
$q$. In principle, one can consider the entire spectrum of the Renyi
entropy and the Renyi divergence. However keeping in mind the finite
and discrete nature of the galaxy samples, it would be sufficient to
consider only the first few $q$ values in the analysis. It may be
noted that considering $q$ values upto 5 instead of 10 does not
significantly change the conclusions of the present analysis.

The results of this analysis agree well with our previous analyses of
the SDSS main galaxy sample and the LRG sample using Shannon entropy
\citep{pandeysarkar15, pandeysarkar16}. The filaments are known to be
the largest known coherent features present in the galaxy
distribution. Earlier studies with SDSS show that they can extend upto
length scales of $100-130 \hmpc$ \citep{pandey05, pandey11}. The
filaments longer than these length scales may be present in the data
but they are not statistically significant and arise purely from
chance alignments. The results of the present analysis are consistent
with these findings.

Our study indicates that if the galaxy samples are not large enough
then the study of homogeneity using the number count based methods
would underestimate the scale of homogeneity. This is caused by the
suppression of the real inhomogeneities by the overlapping bias. The
effects of the overlapping bias can be reduced by considering galaxy
samples with larger size. However, we need to take into account the
evolutionary effects for very large survey volume. The anti-Copernican
models may appear homogeneous in such case. But such models can be
also constrained from other observations such as SNe, CMB and BAO
\citep{zibin, clifton, biswas, chris}. It is also difficult to
construct a precisely volume limited sample from the observational
data due to issues such as the galaxy evolution, the K-corrections,
the angular selection functions and the fibre collisions. Such
limitation can introduce spurious signals of inhomogeneity even in a
homogeneous distribution.

We analyze the SDSS galaxy distribution using independent voxels to
avoid the contamination caused by the overlapping bias. The analysis
with the independent voxels is more sensitive to the presence of
inhomogeneities and clearly shows the absence of homogeneity in the
smaller galaxy samples. Combining the two analyses using the
overlapping spheres and the independent voxels, we identify a
transition to homogeneity in the SDSS galaxy distribution on scales
beyond $140 \hmpc$. The results of these analyses thus reaffirms the
validity of the assumption of cosmic homogeneity.

\section*{Acknowledgments}
We sincerely thank the anonymous reviewer for providing valuable
comments and suggestions that helped us to improve the manuscript. BP
acknowledges the financial support from SERB, DST, Government of India
through the project CRG/2019/001110. BP would also like to acknowledge
IUCAA, Pune for providing support through associateship programme. SS
acknowledges IISER, Tirupati for providing support through a
postdoctoral fellowship.

\end{document}